\documentclass[10pt,conference]{IEEEtran}
\IEEEoverridecommandlockouts

\usepackage{amsmath}
\usepackage{amssymb}
\usepackage{graphicx}
\usepackage{algorithm}
\usepackage{algorithmic}
\usepackage{multirow}
\usepackage{booktabs}
\usepackage{colortbl}
\usepackage{tcolorbox}
\usepackage{url}

\newcommand{\find}[1]{
\begin{tcolorbox}[leftrule=0.5mm,rightrule=0.5mm, toprule=0.5mm,bottomrule=0.5mm,left=2pt,right=2pt,top=2pt,bottom=2pt]
\itshape #1
\end{tcolorbox}
}

\begin{document}

\title{An Iterative Test-and-Repair Framework for Competitive Code Generation}

\author{
\IEEEauthorblockN{
Lingxiao Tang$^{1}$,
Muyang Ye$^{1}$,
Zhaoyang Chu$^{2}$,
Xiaoxue Ren$^{1}$,
Zhongxin Liu$^{1}$,
Lingfeng Bao$^{1,*}$,
He Ye$^{2}$
}
\IEEEauthorblockA{
$^{1}$Zhejiang University\\
$^{2}$University College London\\
\{12421037, yemyuyang, xxren, liu\_zx, lingfengbao\}@zju.edu.cn\\
\{zhaoyang.chu.25, he.ye\}@ucl.ac.uk
}
\thanks{$^{*}$Corresponding author.}
}

\maketitle

\begin{abstract}
Large language models (LLMs) have made remarkable progress in code generation, but competitive programming remains a challenge. 
Recent training-based methods have improved code generation by using reinforcement learning (RL) with execution feedback. The more recent framework CURE further incorporates test generation into the training process, jointly training a Coder and a Tester within a single model. At inference time, the Coder generates many candidate programs, and the Tester generates tests from the problem description. The candidate that passes the largest number of generated tests is selected as the final answer.
However, CURE has two critical limitations. First, the Tester never reads any candidate code, so its tests often fail to expose implementation-specific bugs. Second, the Coder generates every candidate from scratch and never learns to fix a buggy program based on a failed test.
To address these limitations, we propose \textsc{FixAudit}, which approaches competitive code generation from a new perspective: starting from a single initial candidate, it iteratively improves the candidate through a targeted test-and-repair debugging cycle. The framework trains one shared model with two specialized roles through four stages: the Fixer, which repairs the current candidate based on a failing test, and the Auditor, which reads the candidate code to generate new tests that expose its remaining bugs.

We evaluate \textsc{FixAudit} on four benchmarks: APPS, CodeContests, xCodeEval, and LiveCodeBench.
Applied to a 7B model, \textsc{FixAudit} achieves 34.66\% average Pass@1 and 60.28\% average AvgPassRatio, outperforming the larger 32B zero-shot baseline by 37.7\% and 48.1\%, respectively.
Compared with strong baselines built on the same 7B base model, it improves average Pass@1 by 40.2\% to 49.9\% and average AvgPassRatio by 14.4\% to 23.8\%.
Further analysis shows that \textsc{FixAudit} achieves stronger performance with fewer iterations, demonstrating both the effectiveness and efficiency of the approach.
\end{abstract}

\begin{IEEEkeywords}
code generation, reinforcement learning,
large language models
\end{IEEEkeywords}

\section{Introduction}
Large language models (LLMs) have made remarkable progress in code
generation~\cite{program_synthesis_llm,codex_eval,deepseekcoder,qwen25coder}.
However, competitive programming remains a persistent
challenge~\cite{apps,alphacode,xcodeeval,livecodebench,rigorous_eval}.
These problems have strict logical constraints. A model can easily generate a 
program that passes some test examples, but still fails to pass all test sets 
that determine true correctness~\cite{apps,alphacode,rigorous_eval}. 

Reinforcement learning (RL) has emerged as a natural approach for improving code generation by using execution feedback as reward signals.
Early methods such as CodeRL~\cite{le2022coderl}, PPOCoder~\cite{shojaee2023execution}, and RLTF~\cite{liu2023rltf} formulate code generation as policy optimization and use unit test results to guide training.
StepCoder~\cite{dou2024stepcoder} and RLEF~\cite{gehring2024rlef} further improve training stability through curriculum design and multi-turn feedback.
However, these methods focus solely on training the code generation policy and do not train a test generation model.
At inference time, they can only rely on the fixed public tests to evaluate candidates and cannot perform test-time scaling~\cite{muennighoff2025s1,sstar} by generating additional tests to select among multiple solutions.

The state-of-the-art training-based framework CURE~\cite{cure} addresses this limitation by using RL to jointly train a single model to serve as both a Coder and a Tester.
At inference time, the Coder generates a pool of candidate programs, and the Tester generates a set of test cases based solely on the problem description.
The candidate that passes the most of the generated tests is selected as the final answer.
This generate-and-select strategy realizes test-time scaling: generating more candidate programs and more tests together leads to a higher chance of producing and identifying a correct solution.

However, CURE still suffers from two critical limitations.
\textbf{First, the Tester generates tests blindly without analyzing any candidate code.}
Because the Tester only receives the problem description as input and never sees the candidate programs, it produces generic tests that frequently fail to expose hidden, implementation-specific bugs in partially correct code.
\textbf{Second, CURE has no repair mechanism.}
The Coder generates new candidates from scratch each time and relies on the Tester to select the best one.
It never learns to fix a buggy program based on a failed test case.
If none of the candidates in the pool happen to be correct, no amount of test-based selection can produce the right answer.

To overcome these limitations, we propose a different perspective.
Rather than generating many candidates from scratch and selecting among them, we start from a single initial candidate produced by the base model and iteratively improve it through \textbf{a continuous, targeted test-and-repair cycle.} We introduce \textsc{FixAudit}, a new framework that jointly trains one shared model with two roles: the Fixer, which repairs the current candidate, and the Auditor, which generates tests to expose its remaining bugs.

However, neither task is easy for LLMs. Targeted repair requires the model to trace program execution, pinpoint the exact fault, and modify only the faulty logic without breaking correct parts. Targeted test generation requires the model to simulate how the candidate code behaves on specific inputs and identify inputs that trigger its bugs. Both tasks demand strong execution reasoning ability, a capability that current LLMs lack~\cite{chen2025reasoning,gu2024cruxeval,semcoder}. We therefore design a four-stage training pipeline that progressively builds these capabilities:
\textcircled{\raisebox{-0.9pt}{1}} \textbf{Stage~A} equips the model with the execution reasoning ability through supervised fine-tuning. It trains the model both to predict the output of a program on a given input and to derive the correct output from the problem specification alone. The first ability helps the Fixer understand why the current program fails, while the second helps the Auditor construct valid test cases.
\textcircled{\raisebox{-0.9pt}{2}} \textbf{Stage~B} trains the Fixer via RL to repair initial buggy programs using provided failing tests, producing an improved candidate. However, this repaired candidate may still contain latent bugs that the existing tests do not cover.
\textcircled{\raisebox{-0.9pt}{3}} \textbf{Stage~C} therefore trains the Auditor via RL to actively find these hidden bugs. Unlike CURE's Tester, the Auditor reads the repaired candidate alongside the specification, allowing it to generate tests that specifically target the program's remaining bugs.
\textcircled{\raisebox{-0.9pt}{4}} \textbf{Stage~D} closes the loop by retraining the Fixer on these newly exposed failures, further refining the candidate.

We evaluate our framework on four competitive programming benchmarks: APPS, CodeContests, xCodeEval, and LiveCodeBench.
After training and test-time scaling, \textsc{FixAudit} built on Qwen2.5-Coder-7B-Instruct achieves 34.66\% average Pass@1 and 60.28\% average AvgPassRatio.
Within the same model family, it relatively improves average Pass@1 by 37.7\% and average AvgPassRatio by 48.1\% over the larger Qwen2.5-Coder-32B-Instruct used in the zero-shot setting.
Compared with strong baselines built on the same 7B base model~\cite{specine,cure}, \textsc{FixAudit} relatively improves average Pass@1 by 40.2\% to 49.9\% and average AvgPassRatio by 14.4\% to 23.8\%.
Moreover, \textsc{FixAudit} reaches stronger performance with fewer iterations, 
indicating that our approach is both effective and efficient.

In summary, the main contributions of this paper are:
\begin{itemize}
    \item We propose a new perspective for competitive code generation by treating it as a continuous, targeted test-and-repair debugging process centered on the current candidate solution.
    \item We introduce \textsc{FixAudit}, a four-stage training framework for competitive code generation. It first builds the execution reasoning ability through supervised fine-tuning, and then uses RL to train a Fixer for targeted repair and an Auditor for bug-revealing test generation.
    \item We evaluate \textsc{FixAudit} on APPS, CodeContests, xCodeEval, and LiveCodeBench. The results show that our method achieves the strongest overall performance, surpassing larger zero-shot models as well as strong framework baselines.
\end{itemize}

\section{Motivation Example} \label{sec:motivation_example}
\begin{figure}[htbp]
    \centering
    \includegraphics[width=0.85\columnwidth]{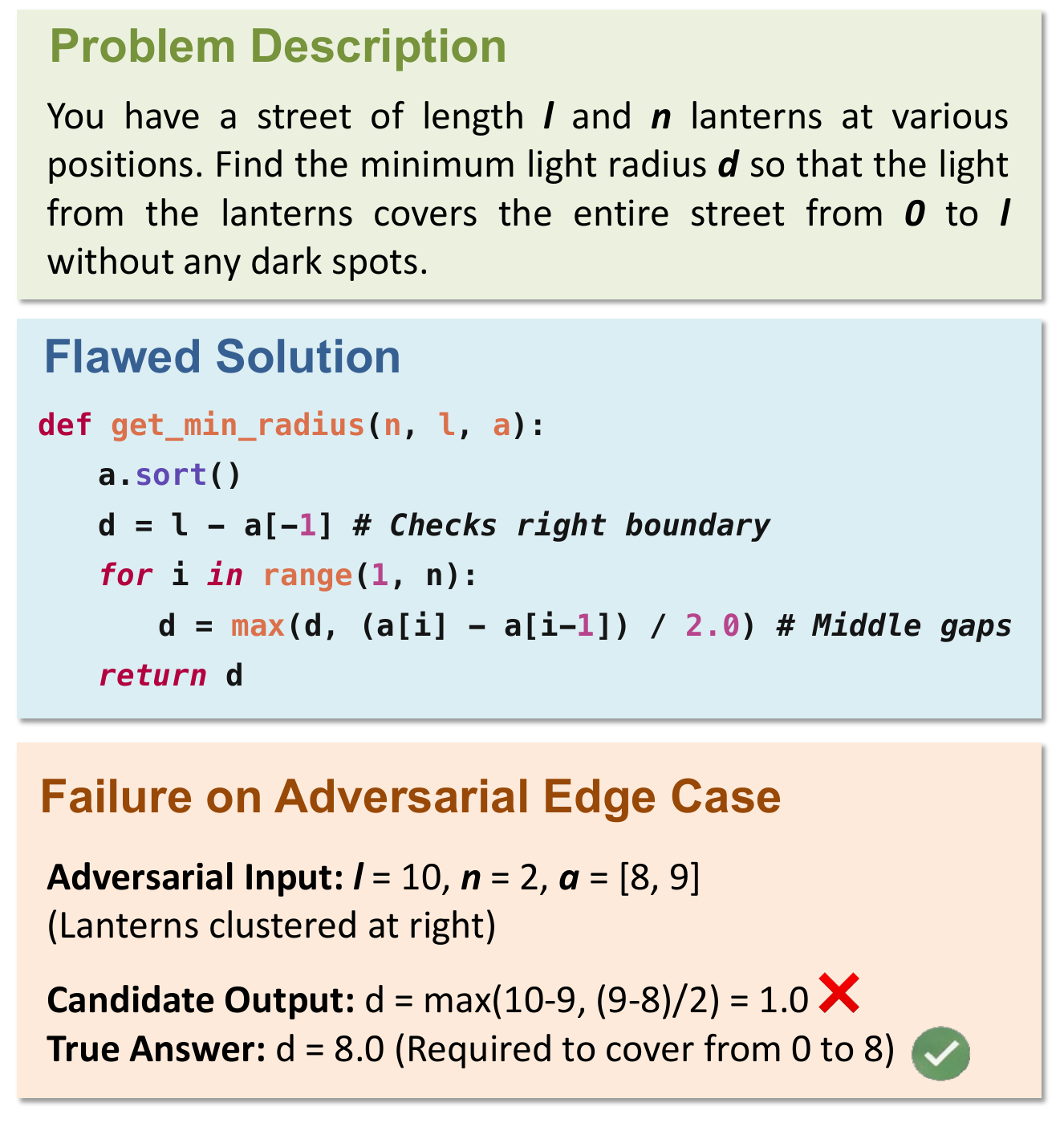}
    \caption{The motivation example}
    \label{fig:motivation}
\end{figure}
This section uses a classic algorithmic problem, ``Street Lanterns'', to illustrate the limitations of the CURE framework and to motivate our test-and-repair approach.

As shown at the top of Figure~\ref{fig:motivation}, the task requires finding the minimum radius $d$ to illuminate an entire street of length $l$ using $n$ lanterns.
A fully correct solution must cover three areas: the left boundary, the middle gaps between lanterns, and the right boundary.
When applying CURE to this problem, we observe that it fails to produce a correct solution.
The candidate programs generated by its Coder frequently contain the flawed logic shown in the middle of Figure~\ref{fig:motivation}.
This deceptive candidate correctly sorts the array, calculates the rightmost boundary gap ($l - a[-1]$), and iterates to find the maximum middle gap ($(a_i - a_{i-1})/2.0$).
Because this partial logic is sufficient to pass all provided public tests, the candidate appears to be correct.
However, there is a severe logic flaw hidden in this code: it completely forgets to check the distance from the start of the street to the first lantern ($a_0$).
As illustrated by the adversarial test case at the bottom of Figure~\ref{fig:motivation}, if all lanterns are placed far to the right, this missing check causes the program to leave the entire left side of the street unlit.
CURE fails to correct this candidate due to its two fundamental limitations:

\begin{itemize}
    \item \textbf{The Tester generates tests without reading the candidate code.}
    CURE's Tester generates test cases based only on the problem description. It never sees the candidate code, so it does not know that the \texttt{a[0]} check is missing.
    Without this information, the Tester is unlikely to construct the specific kind of input needed to trigger this bug, such as placing all lanterns far to the right.
    As a result, the generated tests fail to distinguish this flawed candidate from a correct one, and CURE selects it as the final answer.

    \item \textbf{The Coder has no repair mechanism.}
    Even if a test happened to expose this bug, CURE cannot fix the candidate. Its Coder generates every candidate from scratch and never learns to repair an existing program based on a failing test.
    The only option is to generate more new candidates, hoping that one of them does not contain the same mistake.
    Meanwhile, the correct logic already present in this candidate, including the sorting, the right-boundary check, and the middle-gap calculation, is entirely discarded.
\end{itemize}

This example points to two capabilities that are necessary for solving such problems.
First, \emph{test generation should read the candidate code, not just the problem description}.
If a test generator examines the candidate and notices that \texttt{a[0]} is never used, it can intentionally place all lanterns far from the left boundary to trigger the bug.
Without examining the code, it is much less likely to construct such a targeted input.
Second, \emph{repair should fix the bug in place rather than regenerate the entire solution from scratch}.
The candidate already contains correct logic for sorting, the right boundary, and the middle gaps.
A good repair strategy should preserve these correct parts and only add the missing left-boundary check, instead of discarding everything and risking the same mistake again.
This observation motivates us to design \textsc{FixAudit} as a targeted test-and-repair cycle centered on the current candidate solution.

\section{Approach}

\begin{figure*}[t]
    \centering
    \includegraphics[width=0.85\textwidth]{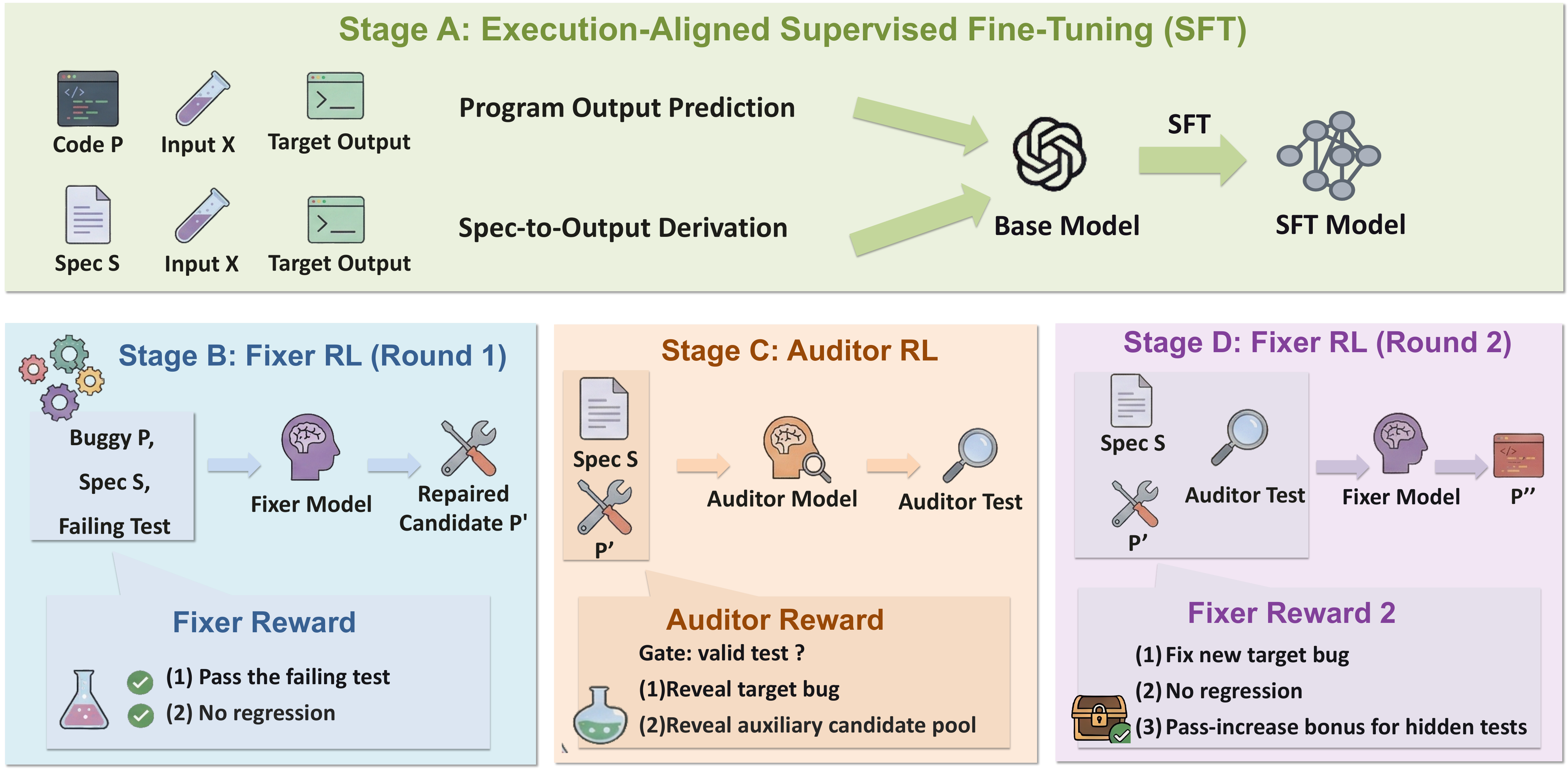}
    \caption{Overview of \textsc{FixAudit}.}
    \label{fig:overview}
\end{figure*}

In this paper, we present \textsc{FixAudit}, an iterative test-and-repair framework that treats the current candidate solution as a target for active debugging.
Our framework defines two roles within a single shared language model: a program repair agent (the Fixer) and a test generation agent (the Auditor).
As illustrated in Figure~\ref{fig:overview}, the approach consists of four training stages.
We first enhance the model's execution reasoning capability through supervised fine-tuning (Stage~A).
We then train the Fixer via reinforcement learning (RL) to repair buggy programs given a failing test while preserving existing correct logic (Stage~B).
Next, we train the Auditor to read the repaired code and generate new test cases (both inputs and expected outputs) that expose its remaining bugs (Stage~C).
Finally, we retrain the Fixer on these Auditor-generated tests to close the loop (Stage~D).

\subsection{Stage A: Execution-Aligned Supervised Fine-Tuning}\label{sec:stage-a}

Targeted debugging requires the model to pinpoint the exact logical flaw within an existing algorithm.
This demands strong execution reasoning: the ability to trace how a program state evolves and predict execution outcomes.
This capability is central to both agents.
For the Fixer, execution reasoning is essential to understand \emph{why} a given test fails: it allows the model to trace the erroneous execution path, pinpoint the fault, and verify that a proposed patch produces the correct behavior.
For the Auditor, execution reasoning enables the agent to anticipate how the target code will behave on different inputs, allowing it to craft test cases that expose specific bugs rather than relying on random guessing.
Because weak execution reasoning would bottleneck both agents, Stage~A builds this foundation \emph{before} any RL training begins.

Formally, we consider problems with a natural-language specification $S$.
Let $P$ denote a candidate program, and let $G$ denote a reference solution assumed to be correct.
We write $\textsc{Run}(P, x)$ for the result of executing program $P$ on an input $x$.
We train the model's execution reasoning through supervised fine-tuning on two complementary tasks.

The first task, \textbf{Program Output Prediction}, trains the model to predict the execution result $\textsc{Run}(P, x)$ of a given program on a given input.
This equips the model with the ability to simulate program behavior, which the Fixer later uses to reason about how a repair changes the program's output.
We construct training samples by pairing sampled inputs with their observed execution outputs.

The second task, \textbf{Specification-to-Output Derivation}, trains the model to derive the expected correct output $\textsc{Run}(G, x)$ for a given input $x$ using only the natural-language specification $S$.
This task is essential for the Auditor's ability to generate valid test cases.
Because the Auditor must propose complete test cases, which include both inputs and expected outputs, this training ensures that it can correctly deduce the required output for any input from the problem description alone.

We mix samples from both tasks uniformly and fine-tune a single model.
By jointly training on both tasks, the model develops an understanding of both code execution and specification semantics, benefiting both the Fixer and the Auditor.
The resulting checkpoint $\pi_0$ serves as the common initialization for all subsequent RL stages.

\subsection{Stage B: Fixer RL (Round-1 Repair)}\label{sec:stage-b}

Building upon the model from Stage~A, Stage~B trains the Fixer via reinforcement learning to repair buggy programs.
Given a problem with specification $S$, the base model first generates an initial candidate program $P$ in a zero-shot manner.
The Fixer then receives $S$, the buggy candidate $P$, and a public test that $P$ fails to pass (denoted $x_f$), and generates a repaired program $P'$.
Unlike CURE's Coder, which generates every candidate from scratch, the Fixer works directly on the existing candidate to produce a targeted fix.

To evaluate repairs, we define $\textsc{Pass}(P', x)$ as an indicator that the repaired program produces the correct output, determined by comparing against the reference solution $G$:
\begin{equation}
    \textsc{Pass}(P', x) = \mathbf{1}\big[\textsc{Run}(P', x) = \textsc{Run}(G, x)\big]. \label{eq:pass}
\end{equation}

The reward design enforces a ``do no harm'' principle to preserve already-correct logic.
We define a regression set $T_{\text{reg}}(P)$ that collects all available tests (both public and hidden) that the original candidate $P$ currently passes.
These tests are used only for computing the reward signal and are never exposed to the model as input.
A regression occurs if the repair breaks any of these previously passing tests:
\begin{equation}\label{eq:regression}
\begin{aligned}
\textsc{Regress}(P \!\to\! P') = \mathbf{1}\Big[{}
  &\exists\,x \in T_{\text{reg}}(P): \\
  &\textsc{Pass}(P,x) \wedge \neg\textsc{Pass}(P',x)
\Big].
\end{aligned}
\end{equation}

We formulate the Round-1 Fixer reward as:
\begin{equation}\label{eq:r-fix-1}
R_{\text{fix}}^{(1)} =
\begin{cases}
0, & \text{if } \textsc{Pass}(P', x_f) = 0, \\
0, & \text{if } \textsc{Regress}(P \to P'), \\
1, & \text{otherwise}.
\end{cases}
\end{equation}

The Fixer receives a reward of 1 only when it both fixes the failing test $x_f$ and does not break any previously passing test.
The zero reward for regressions teaches the model to preserve the already-correct parts of the code and only modify the faulty logic.

\subsection{Stage C: Auditor RL (Bug-Revealing Test Generation)}\label{sec:stage-c}

After Stage~B, the Fixer is trained to repair candidates so that they pass the given failing test without causing regressions.
However, because these repairs are optimized against a fixed set of tests, they may still contain latent bugs that the existing tests do not cover.
Stage~C addresses this by training the Auditor to generate targeted tests that expose these hidden errors.

Unlike CURE, where the Tester generates inputs based only on the problem description, the Auditor in \textsc{FixAudit} reads the candidate code.
Given the specification $S$ and the Fixer's repaired candidate $P'$, the Auditor must propose a complete test case which contains both an input $x$ and its expected output $y$, that breaks $P'$.
Because the Auditor sees the actual code, it can focus on the specific bugs in the current solution rather than generating tests blindly.

To formulate the Auditor's reward, we define two conditions for a valid bug-revealing test.
First, the proposed output $y$ must match the ground-truth output produced by the reference solution $G$.
Second, the test must cause the repaired program to produce a wrong answer:
\begin{align}
    \textsc{Valid}(x, y) &= \mathbf{1}\big[\textsc{Run}(G, x) = y\big], \label{eq:valid} \\
    \textsc{Reveal}(P', x, y)
      &= \mathbf{1}\Big[\textsc{Valid}(x,y) \notag\\
      &\qquad\wedge \big(\textsc{Run}(P',x) \neq y\big)\Big]. \label{eq:reveal}
\end{align}

Rewarding the Auditor only for breaking the target repair $P'$ yields a very sparse signal, because most generated test cases do not happen to trigger edge-case bugs.
To provide denser feedback, we introduce reward shaping over an auxiliary pool $\mathcal{P}_{\text{aux}}$ of $k$ alternative  repairs sampled from the Round-1 Fixer for the same problem.
The Auditor reward for generating a test $(x, y)$ is:

\begin{equation}\label{eq:r-audit}
\begin{aligned}
R_{\text{audit}}(x, y)
={}& \textsc{Valid}(x,y)\Bigg[ \\
&\underbrace{0.5\,\textsc{Reveal}(P',x,y)}_{\text{primary}} \\
&\quad+\underbrace{
0.5\!\sum_{\tilde{P}\in\mathcal{P}_{\text{aux}}}
\!\textsc{Reveal}(\tilde{P},x,y)
}_{\text{auxiliary shaping}}\Bigg].
\end{aligned}
\end{equation}

The outer $\textsc{Valid}(x, y)$ factor gates the entire reward: the Auditor receives no reward unless it produces a valid test case.
The primary term rewards the main objective, namely revealing the bug in the target candidate $P'$.
The auxiliary shaping term further rewards test cases that also reveal bugs in other incorrect repairs from the auxiliary pool $\mathcal{P}_{\text{aux}}$.
Both the primary objective and the auxiliary shaping signal are weighted by $0.5$.
This encourages tests that capture common failure patterns.

\subsection{Stage D: Fixer RL (Round-2 Refinement)}\label{sec:stage-d}

Stage~D closes the loop by retraining the Fixer on the bugs discovered by the Auditor.
In Stage~B, the Fixer was trained on a fixed set of failing tests.
Now, by feeding it the new, targeted failures from the Auditor, we push the Fixer to address deeper logical flaws in the candidate.

For each Round-1 repair $P'$, the trained Auditor generates a bug-revealing test case $(x_{\text{audit}}, y_{\text{audit}})$.
The Round-2 Fixer then takes $P'$ and this new test to produce a further refined repair $P''$.

Because the Auditor is an imperfect agent, its expected output $y_{\text{audit}}$ may occasionally be incorrect.
We therefore do not require the Fixer to pass this specific test.
Instead, we treat it as a hint that directs the Fixer's attention to a potential flaw in its current logic.
The Fixer is rewarded based solely on the ground-truth hidden test suite, so it must judge whether the hint points to a real bug and improve its solution accordingly.

We measure improvement by counting how many additional hidden tests ($T_{\text{hid}}$) the new repair $P''$ passes compared to $P'$:
\begin{equation}\label{eq:delta-hid}
\begin{aligned}
\Delta_{\text{hid}}(P' \to P'')
  = \sum_{x \in T_{\text{hid}}} \max\Big(&0,\\[-1mm]
  &\textsc{Pass}(P'',x)-\textsc{Pass}(P',x)\Big).
\end{aligned}
\end{equation}

The Round-2 Fixer reward is:
\begin{equation}\label{eq:r-fix-2}
R_{\text{fix}}^{(2)} =
\begin{cases}
0, & \text{if } \textsc{Regress}(P' \to P''), \\
\begin{aligned}
  &\min\big(1.0,0.1\,\Delta_{\text{hid}}\big)\\[-1mm]
  &\quad+\textsc{AllPass}(P'')
\end{aligned}
& \text{otherwise},
\end{cases}
\end{equation}
where $\textsc{AllPass}(P'')$ equals 1 only if $P''$ passes every test in $T_{\text{hid}}$.
As in Stage~B, any regression immediately results in zero reward.
Beyond this gate, the reward increases with the number of newly passed hidden tests (up to 1.0), and an additional bonus of $+1.0$ is given for achieving full correctness.
This design encourages the Fixer to fix the underlying logical flaw rather than overfitting to the Auditor's specific test case.

\subsection{Reinforcement Learning with DAPO}\label{sec:dapo}

All RL stages in our framework (Stages~B, C, and~D) are optimized using DAPO (Decoupled Clip and Dynamic Sampling Policy Optimization)~\cite{dapo}.
The core objective function of DAPO is formulated as:
\begin{equation}\label{eq:dapo}
\begin{aligned}
\mathcal{L}_{\text{DAPO}}(\theta)
={}&\mathbb{E}\Bigg[\frac{1}{N}\sum_{i=1}^{N}\hat{A}_i\,
\min\Big( r_i(\theta),\\
&\qquad \text{clip}\big(r_i(\theta),
1-\epsilon_l,1+\epsilon_u\big)\Big)\Bigg].
\end{aligned}
\end{equation}

Here, $N$ outputs are sampled from the old policy for each prompt, and $r_i(\theta) = \pi_\theta(o_i \mid q) / \pi_{\theta_{\text{old}}}(o_i \mid q)$ is the probability ratio between the updated and old policies for the $i$-th output.
The advantage $\hat{A}_i$ is estimated by normalizing the rewards within the sampled group.
The variables $\epsilon_l$ and $\epsilon_u$ are decoupled lower and upper clipping bounds.

DAPO offers two advantages over its predecessors PPO~\cite{ppo} and GRPO~\cite{grpo}.
First, following GRPO, it computes advantages through group-relative normalization rather than maintaining a separate value network, reducing memory overhead during LLM training.
Second, unlike GRPO, which clips upward and downward policy changes equally ($\epsilon_l = \epsilon_u$), DAPO decouples the two bounds and sets $\epsilon_u > \epsilon_l$.
In practice, this means that when the model discovers a rare successful output (e.g., a bug fix that passes all tests, or a test that breaks the candidate), DAPO allows a large policy update to reinforce it.
Conversely, it applies a tighter bound when decreasing the probability of other outputs, preventing the model from prematurely narrowing its search.
This property is well-suited to our framework, where both the Fixer and the Auditor face sparse rewards and need sustained exploration to discover effective repairs and bug-revealing tests.

\section{Experimental Setup}
\subsection{Baselines}
We compare \textsc{FixAudit} against baselines from three categories.

\textbf{Prompt-based and agent-based baselines.}
Recent work has explored prompt-based and agent-based workflows for iterative code refinement~\cite{selfdebugging,self_edit,selfrepair,selfcollaboration,agentcoder,paircoder}.
These methods use execution feedback or role specialization to improve generated code, but rely on prompting alone without training the model's underlying capabilities.
We select two representative baselines from this category.
\textbf{Self-Repair}~\cite{selfrepair} is a prompt-based repair method that directly puts the failed test's execution feedback into the prompt and asks the model to fix the buggy code. By comparing with Self-Repair, we demonstrate that training a dedicated Fixer via RL is necessary.
\textbf{Specine}~\cite{specine} is the most recent and strongest-performing agent-based framework. It extracts the model-perceived specification from buggy code, rewrites the prompt to reduce ambiguity, and regenerates a new program from scratch.
To ensure a fair comparison, we instantiate both baselines using the same backbone model as \textsc{FixAudit}, Qwen2.5-Coder-7B-Instruct~\cite{qwen2.5}.

\textbf{Training-based baseline.}
We select \textbf{CURE}~\cite{cure}, the state-of-the-art RL-based framework that jointly trains a Coder and a Tester within a single model.
At inference time, CURE generates a pool of candidate programs and a set of tests, then selects the candidate that passes the most generated tests.
We implement CURE on our exact same training set using the Qwen2.5-Coder-7B-Instruct base model.
To ensure a consistent training methodology, we optimize CURE using the same DAPO algorithm and follow its other original settings.

\textbf{Zero-shot foundation models.}
To provide a broader context, we include larger or well-known models evaluated under a zero-shot setting.
We select \textbf{Qwen2.5-Coder-14B-Instruct} and \textbf{Qwen2.5-Coder-32B-Instruct} from the same model family~\cite{qwen2.5}, as well as \textbf{DeepSeek-Coder-v1.5}~\cite{deepseekcoder} and the commercial API model \textbf{GPT-4o-mini}~\cite{gpt4o}.
These models generate code directly from the problem description and serve as references for whether our training framework on a 7B model can match or outperform larger models and commercial alternatives.

\subsection{Evaluation Benchmarks}
We evaluate \textsc{FixAudit} on four competition-level programming benchmarks.
Following recent studies like Specine~\cite{specine}, we exclude basic datasets such as HumanEval~\cite{codex_eval} and MBPP~\cite{program_synthesis_llm}.
Modern LLMs already achieve near-perfect Pass@1 on these basic datasets under zero-shot settings.
Instead, we focus on problems with complex logical constraints to better assess true correctness.
For APPS~\cite{apps}, CodeContests~\cite{alphacode}, and xCodeEval~\cite{xcodeeval}, we use the exact same datasets and sampled subsets as Specine; we additionally evaluate every method on the LiveCodeBench benchmark~\cite{livecodebench}.
\textbf{APPS}~\cite{apps} contains problems from various competitive programming platforms.
We use the 300 problems sampled from its test set based on difficulty distribution.
\textbf{CodeContests}~\cite{alphacode} is a benchmark from Google DeepMind for highly challenging programming tasks.
We use all 165 test problems with the extended version that provides $\sim$190 additional tests per problem.
\textbf{xCodeEval}~\cite{xcodeeval} is a large-scale execution-based benchmark collected from Codeforces.
We similarly use its 300 sampled test problems to match the evaluation scale.
\textbf{LiveCodeBench v6}~\cite{livecodebench} is a recent contamination-resistant benchmark containing competitive programming problems released over time.
Following Specine, we use two complementary evaluation metrics.
The primary metric is \textbf{Pass@1}, which considers a program correct only if it passes all hidden private tests.
To measure partial correctness, we also report the \textbf{Average Pass Ratio (AvgPassRatio)}.
This metric calculates the ratio of passed private test cases for each problem.
\subsection{Training Details}

\textbf{Training Data Construction.}
We construct our foundational training set based on the TACO dataset~\cite{taco}.
We filter the dataset to retain only problems containing $\ge 20$ test cases, yielding $7,463$ problems.
To rigorously prevent data leakage between our training set and downstream evaluation benchmarks, we apply strict $N$-gram filtering~\cite{codex_eval, brown2020language}, obtaining a final, clean corpus of $6,981$ unique problems.

\textbf{Stage A (SFT) Data Collection.}
For the Execution-Aligned Supervised Fine-Tuning, we construct a dataset containing two tasks.
We utilize QwQ-32B~\cite{qwq32b} as the teacher model.
For Program Output Prediction, Qwen2.5-Coder-7B-Instruct generates a candidate solution, and the teacher predicts the output for a randomly selected test input.
For Specification-to-Output Derivation, the teacher predicts the correct output based solely on the problem specification and a random input.
To ensure diversity, a different test input is used during each iteration.
From an initial pool of $20,000$ samples per task, rejection sampling against ground-truth outputs yielded $17,243$ and $14,946$ high-quality samples, respectively.

\textbf{Training Configurations.}
Both the SFT and the adversarial reinforcement learning (RL) stages are implemented using the open-source \textbf{veRL}~\cite{verl} framework.
For the SFT stage, we fine-tune the model with a learning rate of $1 \times 10^{-5}$.
For the RL stages (Stages B through D), we utilize all $6,981$ problems from the filtered dataset.
The environment uses the dataset's ground-truth reference solutions and test cases to validate generated outputs and compute reward signals.
We use DAPO~\cite{dapo} as the RL algorithm, with GRPO~\cite{grpo} as the advantage estimator.
During RL training, we set the actor learning rate to $1 \times 10^{-6}$.
The maximum prompt length is set to $2,048$ tokens, and the maximum response length is $8,192$ tokens.
All training experiments are conducted on a cluster of $8 \times$ NVIDIA A100 (80GB) GPUs.
The RL training runs for a total of $100$ steps for each stage.

\subsection{Inference Details}
To ensure a fair comparison, we restrict all frameworks to the same inference budget of 20 LLM invocations per problem.
For Specine, following its default setting, we allow a maximum of 10 iterations. Each iteration requires calling an Aligner agent and a Coder agent, which results in up to 20 LLM calls.
For CURE, which relies on parallel sampling rather than sequential interaction, we configure it to generate exactly 10 candidate solutions and 10 unit tests per problem, summing to 20 LLM calls.
For \textsc{FixAudit}, a single full repair cycle consists of four sequential steps: generating an initial solution with the base model, applying the Round-1 Fixer to address public test failures, using the Auditor to generate adversarial tests, and finally applying the Round-2 Fixer for refinement.
Under the budget constraint, we allow our framework to iterate through this test-and-repair cycle up to 5 times (yielding $4 \times 5 = 20$ LLM invocations).
To select the final code submission from these iterations, we first filter for candidates that successfully pass all public tests.
If multiple candidates pass, we select the one that passes the highest number of valid test cases generated by the Auditor during the process.
By keeping the invocation budget equal across all frameworks, we ensure that performance gains of \textsc{FixAudit} come from our training framework rather than extra compute at inference time.
For all models and baselines during inference, we standardize the generation hyperparameters by setting the temperature to $1.0$ and top-$p$ to $1.0$.

\section{Experimental Results}
\subsection{RQ1: How effective is \textsc{FixAudit} compared to baselines?}

\begin{table*}[t]
\centering
\caption{Effectiveness comparison in terms of Pass@1 ($\uparrow$) and AvgPassRatio ($\uparrow$). APR is short for AvgPassRatio, and Average is the macro-average across four benchmarks. Best results are highlighted in bold.}
\label{tab:main_results}
\resizebox{0.9\textwidth}{!}{
\begin{tabular}{cl *{10}{c}}
\toprule
\multirow{2}{*}{\textbf{LLM}} & \multirow{2}{*}{\textbf{Technique}} & \multicolumn{2}{c}{\textbf{APPS}} & \multicolumn{2}{c}{\textbf{CodeContests}} & \multicolumn{2}{c}{\textbf{xCodeEval}} & \multicolumn{2}{c}{\textbf{LiveCodeBench}} & \multicolumn{2}{c}{\textbf{Average}} \\
\cmidrule{3-4} \cmidrule{5-6} \cmidrule{7-8} \cmidrule{9-10} \cmidrule{11-12}
& & \textbf{Pass@1} & \textbf{APR} & \textbf{Pass@1} & \textbf{APR} & \textbf{Pass@1} & \textbf{APR} & \textbf{Pass@1} & \textbf{APR} & \textbf{Pass@1} & \textbf{APR} \\
\midrule
DeepSeek-Coder-v1.5 & Zero-shot & 18.00\% & 34.25\% & 2.42\% & 13.02\% & 11.00\% & 24.77\% & 15.18\% & 33.50\% & 11.65\% & 26.39\% \\
GPT-4o-mini & Zero-shot & 33.67\% & 43.61\% & 4.85\% & 17.96\% & 23.67\% & 32.82\% & 24.11\% & 39.00\% & 21.58\% & 33.35\% \\
\midrule
\multirow{3}{*}{Qwen2.5-Coder} & Zero-shot (7B) & 19.67\% & 36.66\% & 5.45\% & 18.64\% & 10.67\% & 27.63\% & 17.86\% & 37.13\% & 13.41\% & 30.02\% \\
& Zero-shot (14B) & 28.70\% & 47.10\% & 6.00\% & 18.20\% & 19.00\% & 42.00\% & 19.64\% & 26.48\% & 18.34\% & 33.45\% \\
& Zero-shot (32B) & 35.70\% & 52.80\% & 11.00\% & 20.70\% & 26.30\% & 47.70\% & 27.68\% & 41.61\% & 25.17\% & 40.70\% \\
\midrule
\multirow{4}{*}{Qwen2.5-Coder-7B} & Self-Repair & 21.67\% & 37.12\% & 6.06\% & 15.13\% & 12.00\% & 24.17\% & 18.75\% & 38.50\% & 14.62\% & 28.73\% \\
& Specine & 37.70\% & 60.60\% & 10.30\% & 34.10\% & 18.70\% & 42.10\% & 30.36\% & 57.98\% & 24.34\% & 48.70\% \\
& CURE & 37.30\% & 65.20\% & 11.90\% & 38.70\% & 18.30\% & 55.20\% & 25.00\% & 51.65\% & 23.13\% & 52.69\% \\
& \textit{\textbf{\textsc{FixAudit}}} & \textbf{46.67\%} & \textbf{67.65\%} & \textbf{30.91\%} & \textbf{50.60\%} & \textbf{28.00\%} & \textbf{55.94\%} & \textbf{33.04\%} & \textbf{66.91\%} & \textbf{34.66\%} & \textbf{60.28\%} \\
\bottomrule
\end{tabular}
}
\end{table*}

Table~\ref{tab:main_results} presents the overall effectiveness comparison.
\textsc{FixAudit} achieves the best average Pass@1 (34.66\%) and AvgPassRatio (60.28\%) among all compared methods.
This shows that \textsc{FixAudit} not only improves the chance of producing fully correct solutions, but also increases the overall correctness of partially correct solutions.

\textbf{Comparison with larger zero-shot models.}
Although \textsc{FixAudit} uses only a 7B backbone, it outperforms substantially larger zero-shot models.
Compared with Qwen2.5-Coder-32B, \textsc{FixAudit} relatively improves average Pass@1 by 37.7\% and AvgPassRatio by 48.1\%.
It also relatively outperforms GPT-4o-mini by 60.6\% in average Pass@1.

\textbf{Comparison with the prompt-based repair baseline.}
Self-Repair uses failed test execution feedback in the prompt, but it does not perform dedicated training.
Its average Pass@1 is 14.62\%, only slightly higher than the 7B zero-shot baseline of 13.41\%.
This indicates that prompting alone provides limited repair ability, because the model may still fail to locate the root cause or make a consistent fix from feedback.
In contrast, \textsc{FixAudit} trains the Fixer through RL, leading to a much higher average Pass@1 of 34.66\%.
This confirms the importance of repair-oriented training.

\textbf{Comparison with framework baselines on the same 7B model.}
Under the same 7B backbone, \textsc{FixAudit} clearly outperforms Specine and CURE.
It relatively improves average Pass@1 by 40.2\% and 49.9\%, and AvgPassRatio by 23.8\% and 14.4\%, respectively.
The largest advantage appears on CodeContests, where \textsc{FixAudit} achieves 30.91\% Pass@1, about $3.0\times$ Specine (10.30\%) and $2.6\times$ CURE (11.90\%).
This benchmark contains more complex algorithmic logic, so public tests are often insufficient to expose hidden bugs.
The code-aware Auditor helps by generating targeted tests for candidate-specific errors, making \textsc{FixAudit} more effective at distinguishing and repairing partially correct solutions.

\find{
\textbf{Answer to RQ1:}
    \textsc{FixAudit} achieves the best overall performance.
    It relatively surpasses the larger Qwen2.5-Coder-32B by 37.7\% in average Pass@1 and 48.1\% in average AvgPassRatio.
    Compared with framework baselines on the same 7B model, it relatively improves average Pass@1 by 40.2\% over Specine and 49.9\% over CURE.
}

\subsection{RQ2: How does each component of \textsc{FixAudit} contribute to performance?}

To understand the contribution of each component, we design six ablation variants.
\textbf{w/o Stage A (SFT)} removes execution-aligned supervised fine-tuning and starts RL directly from the base model, testing whether execution-reasoning initialization is necessary.
\textbf{Stage B Only} retains only Round-1 Fixer RL and repeats it for 10 iterations so that its candidate count matches the full pipeline.
\textbf{w/o Auditor-based Test Selection} keeps all training stages but removes the inference-time signal used to select among candidates that pass the public tests.
\textbf{Blind Auditor (w/o Candidate Code)} provides the Auditor with only the problem description, testing whether candidate-code access enables targeted test generation.
\textbf{w/o Auxiliary Shaping (Stage~C)} removes the auxiliary candidate-pool term from the Auditor reward in Eq.~\ref{eq:r-audit}.
\textbf{Forced Auditor Test (Stage~D)} requires the Round-2 Fixer to pass the generated test instead of treating that test as a hint.

\begin{table*}[t]
\centering
\caption{Ablation study of key components in \textsc{FixAudit}. APR is short for AvgPassRatio, and Average is the macro-average across four benchmarks. Best results are highlighted in bold.}
\label{tab:ablation_results}
\resizebox{0.9\textwidth}{!}{
\begin{tabular}{l *{10}{c}}
\toprule
\multirow{2}{*}{\textbf{Variant}} & \multicolumn{2}{c}{\textbf{APPS}} & \multicolumn{2}{c}{\textbf{CodeContests}} & \multicolumn{2}{c}{\textbf{xCodeEval}} & \multicolumn{2}{c}{\textbf{LiveCodeBench}} & \multicolumn{2}{c}{\textbf{Average}} \\
\cmidrule{2-3} \cmidrule{4-5} \cmidrule{6-7} \cmidrule{8-9} \cmidrule{10-11}
& \textbf{Pass@1} & \textbf{APR} & \textbf{Pass@1} & \textbf{APR} & \textbf{Pass@1} & \textbf{APR} & \textbf{Pass@1} & \textbf{APR} & \textbf{Pass@1} & \textbf{APR} \\
\midrule
w/o Stage A (SFT) & 36.30\% & 63.30\% & 21.10\% & 45.60\% & 22.70\% & 55.40\% & 30.36\% & 62.00\% & 27.62\% & 56.58\% \\
Stage B Only & 34.00\% & 60.30\% & 17.00\% & 43.60\% & 21.00\% & 52.80\% & \textbf{33.04\%} & 65.07\% & 26.26\% & 55.44\% \\
w/o Auditor-based Test Selection & 44.30\% & 67.20\% & 25.90\% & 43.18\% & 26.60\% & 55.70\% & 32.14\% & 65.21\% & 32.24\% & 57.82\% \\
Blind Auditor (w/o Candidate Code) & 40.30\% & 66.20\% & 24.20\% & 42.15\% & 23.50\% & 54.90\% & 32.14\% & 64.50\% & 30.04\% & 56.94\% \\
w/o Auxiliary Shaping (Stage~C) & 43.20\% & 67.15\% & 28.60\% & 46.42\% & 24.50\% & 55.50\% & \textbf{33.04\%} & 65.80\% & 32.34\% & 58.72\% \\
Forced Auditor Test (Stage~D) & 42.20\% & 66.30\% & 26.30\% & 45.14\% & 23.65\% & 55.20\% & 32.14\% & 65.11\% & 31.07\% & 57.94\% \\
\textbf{\textsc{FixAudit}} & \textbf{46.67\%} & \textbf{67.65\%} & \textbf{30.91\%} & \textbf{50.60\%} & \textbf{28.00\%} & \textbf{55.94\%} & \textbf{33.04\%} & \textbf{66.91\%} & \textbf{34.66\%} & \textbf{60.28\%} \\
\bottomrule
\end{tabular}
}
\end{table*}

Table~\ref{tab:ablation_results} shows that all variants perform worse than the full \textsc{FixAudit} model on average, indicating that each component contributes to the final performance.

\textbf{Stage~A and the Auditor loop are important.}
Removing Stage~A reduces average Pass@1 from 34.66\% to 27.62\%, showing that execution-aligned initialization helps the model understand buggy behavior.
Using only Stage~B causes the largest drop, reducing average Pass@1 to 26.26\%.
This indicates that public tests alone are insufficient, and the Auditor loop provides useful additional tests to expose hidden bugs.

\textbf{Code-aware auditing improves test generation and selection.}
Removing Auditor-based selection lowers average Pass@1 to 32.24\%, showing that generated tests help select better candidates that already pass public tests.
The Blind Auditor further reduces average Pass@1 to 30.04\%, especially on CodeContests.
This confirms that access to candidate code helps the Auditor generate more targeted tests for implementation-specific bugs.
Removing auxiliary shaping also decreases average Pass@1 to 32.34\%.
This indicates that auxiliary shaping provides useful training signals when bug-revealing tests are still sparse.

\textbf{Generated tests are better used as hints than as hard constraints.}
Forcing the Fixer to pass Auditor-generated tests reduces average Pass@1 to 31.07\%.
This suggests that generated tests should guide the Fixer rather than strictly constrain it.

\find{
\textbf{Answer to RQ2:}
Each component improves \textsc{FixAudit}.
The Auditor loop provides the largest gain, and code-aware auditing is important for generating targeted tests.
Other components provide complementary benefits.}

\subsection{RQ3: How do \textsc{FixAudit} and the baselines perform as the number of iterations increases?}

In this experiment, we track how performance changes as each method performs more iterations.
For \textsc{FixAudit}, one full iteration produces two candidate solutions: one after the Round-1 Fixer and one after the Round-2 Fixer.
To align the comparison by the number of generated candidates, we match one \textsc{FixAudit} iteration with every two baseline iterations.

\begin{figure*}[t]
\centering
\includegraphics[width=\textwidth]{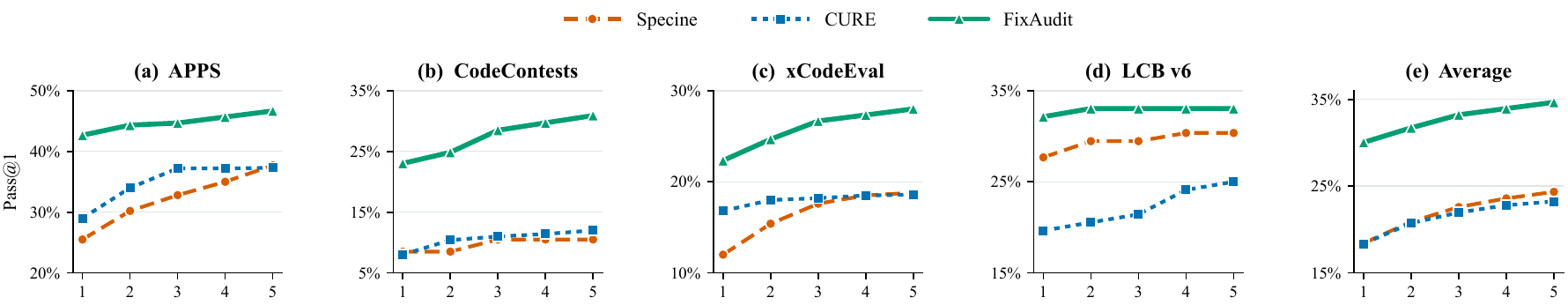}
\caption{Pass@1 performance as the number of iterations increases.}
\label{fig:iteration_results}
\vspace{-3mm}
\end{figure*}

Figure~\ref{fig:iteration_results} shows that \textsc{FixAudit} consistently outperforms both baselines across all aligned iterations.
The advantage is visible from the very first iteration and remains stable throughout.
Notably, after just one complete test-and-repair cycle, \textsc{FixAudit} already surpasses the \emph{final} performance of both Specine and CURE on all four benchmarks.
This indicates that \textsc{FixAudit} is not only more effective but also more efficient at converting iterations into performance gains.

The gap is largest on APPS and CodeContests.
After five aligned iterations, \textsc{FixAudit} reaches 46.67\% on APPS, versus 37.70\% for Specine and 37.30\% for CURE.
On CodeContests, it reaches 30.91\%, versus 10.50\% and 12.00\%, respectively.
It also leads on xCodeEval (28.00\% vs.\ 18.80\% and 18.60\%) and LiveCodeBench (33.04\% vs.\ 30.36\% and 25.00\%).
On average, \textsc{FixAudit} rises from 30.04\% to 34.66\%, while Specine and CURE grow from 18.42\%/18.36\% to 24.34\%/23.23\%.

We attribute this efficiency gap to how each framework uses additional iterations.
Specine rewrites the specification and regenerates the program in each iteration, while CURE samples candidates mostly independently and generates tests without reading candidate code.
As a result, later iterations provide limited guidance from previous attempts.
In contrast, \textsc{FixAudit} performs incremental refinement: the Auditor exposes a specific bug in the current candidate, and the Fixer directly targets that bug.
This makes each iteration more informative, allowing \textsc{FixAudit} to reach stronger solutions earlier and maintain a clear advantage as iterations continue.

\find{
\textbf{Answer to RQ3:}
\textsc{FixAudit} consistently outperforms Specine and CURE across aligned iterations.
Notably, after only one complete test-and-repair cycle, it already exceeds their final performance on all four benchmarks.
This shows that \textsc{FixAudit} is both more effective and more iteration-efficient.
}

\section{Discussion}
\subsection{Generalization Across Model Families}

To check whether the improvements only come from the Qwen backbone, we also evaluate \textsc{FixAudit} with Meta-Llama-3-8B under the same training and inference settings. As shown in Table~\ref{tab:llama_generalization}, \textsc{FixAudit} achieves the best Pass@1 on all four benchmarks. Its average Pass@1 is 11.20\%, higher than zero-shot (4.55\%), Self-Repair (5.26\%), and Specine (8.11\%). Compared with the strongest baseline, Specine, \textsc{FixAudit} improves average Pass@1 by 38.1\%. These results suggest that the test-and-repair loop is not limited to the Qwen model family. Even with a weaker initial generator, \textsc{FixAudit} can still use generated tests to expose bugs and guide repair.

\begin{table}[t]
\centering
\caption{Pass@1 (\%) with Meta-Llama-3-8B.}
\label{tab:llama_generalization}
\resizebox{0.9\columnwidth}{!}{
\begin{tabular}{lccccc}
\toprule
\textbf{Method} & \textbf{APPS} & \textbf{CodeContests} & \textbf{xCodeEval} & \textbf{LCB v6} & \textbf{Average} \\
\midrule
Zero-shot & 6.00 & 0.61 & 0.00 & 11.61 & 4.55 \\
Self-Repair & 7.00 & 1.21 & 0.33 & 12.50 & 5.26 \\
Specine & 13.00 & 1.82 & 0.67 & 16.96 & 8.11 \\
\textbf{\textsc{FixAudit}} & \textbf{18.00} & \textbf{4.24} & \textbf{2.00} & \textbf{20.54} & \textbf{11.20} \\
\bottomrule
\end{tabular}
}
\end{table}

\subsection{Effectiveness of Auditor-Generated Tests}

We further analyze the quality of the tests generated by the Auditor on APPS and CodeContests. 
We collect all tests generated across five iterations. 
Because some problems do not provide a ground-truth solution, a portion of the generated tests cannot be reliably compared and is therefore removed. 
After this filtering step, 1,039 checkable tests remain on APPS and 541 remain on CodeContests. 
A test is \textbf{valid} if the output generated in the test matches the output produced by the ground-truth solution on the same input. 
A test is \textbf{bug-revealing} if it is valid and the target candidate produces a different output from the ground-truth solution on that input. 
Among valid tests, we further distinguish \textbf{valid and candidate-correct} tests, on which the target candidate also produces the correct output.

\begin{table}[t]
\centering
\caption{Quality analysis of Auditor-generated tests on checkable cases.}
\label{tab:auditor_test_quality}
\resizebox{0.9\columnwidth}{!}{
\begin{tabular}{lcc}
\toprule
\textbf{Metric} & \textbf{APPS} & \textbf{CodeContests} \\
\midrule
\# Checkable Tests & 1039 & 541 \\
Valid & 785 (75.6\%) & 383 (70.8\%) \\
Invalid & 254 (24.4\%) & 158 (29.2\%) \\
Valid and Bug-Revealing & 134 (17.1\% of valid) & 173 (45.2\% of valid) \\
Valid and Candidate-Correct & 651 (82.9\% of valid) & 210 (54.8\% of valid) \\
\bottomrule
\end{tabular}
}
\end{table}

Table~\ref{tab:auditor_test_quality} shows that most checkable tests generated by the Auditor are valid. 
The validity rate reaches 75.6\% on APPS and 70.8\% on CodeContests, which indicates that the Auditor can correctly infer the expected output in most cases. 
The slightly lower validity on CodeContests is also expected, since these problems are more difficult and require stronger reasoning to derive the correct output.

More importantly, a substantial fraction of valid tests are also bug-revealing. 
On APPS, 134 valid tests expose errors in the target candidate, accounting for 17.1\% of all valid tests. 
On CodeContests, this ratio increases to 45.2\%. 
This result suggests that Auditor-generated tests are not merely valid, but also highly targeted. 
The larger ratio on CodeContests further indicates that these tests become even more valuable on harder problems, where candidate solutions contain more residual errors and a correct test is more likely to expose them. 
We do not report the same analysis on xCodeEval and LiveCodeBench because no ground-truth Python3 solution is available for reliable validation.

\subsection{A Case Study of Fixer-Auditor Refinement}

\begin{figure}[t]
    \centering
    \includegraphics[width=0.85\columnwidth]{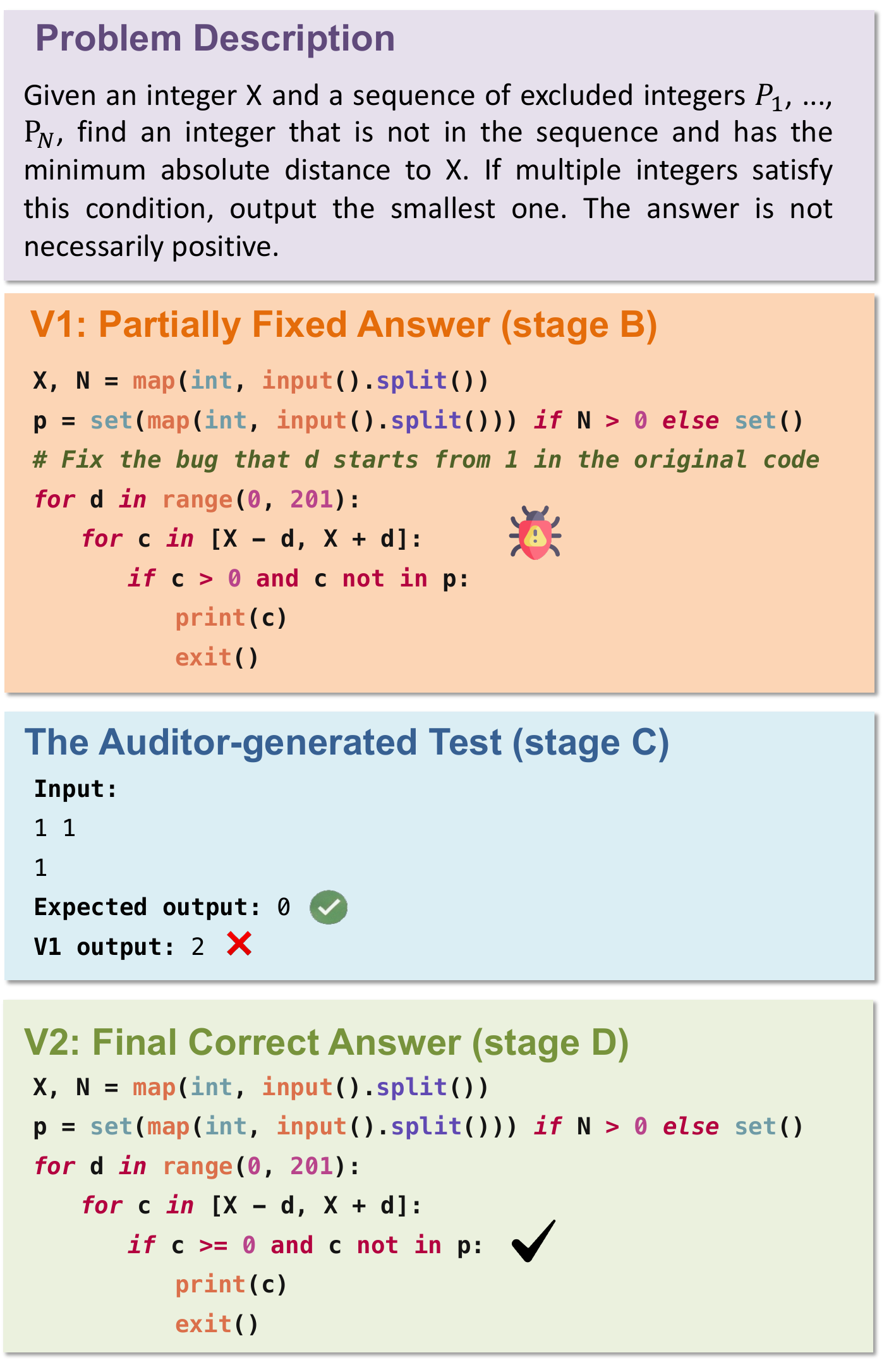}
    \caption{A case study on the problem, ``Nearest Integer Not in Sequence.''}
    \label{fig:discussion_case}
    \vspace{-3mm}
\end{figure}

As shown in Figure~\ref{fig:discussion_case}, the problem, ``Nearest Integer Not in Sequence,'' provides a representative example of how \textsc{FixAudit} benefits from Fixer-Auditor refinement. 
The task asks the model to return the integer that is not in the excluded set and is closest to $X$, breaking ties by choosing the smaller integer. 
Crucially, the answer is not necessarily positive. 
This detail makes the value 0 a valid candidate, which is easy to overlook when the current program already passes all public tests.

In Stage~B, the Fixer first repairs an obvious bug in the initial solution. 
The original program starts its search from distance 1 and therefore fails to consider $X$ itself. 
After observing this failure on the public tests, the Fixer changes the search range to start from 0, which produces the partially fixed solution $V_1$. 
This Stage~B repair directly addresses the public-test failure in the initial solution, and the corresponding fix is described in the comment in the $V_1$ part of Figure~\ref{fig:discussion_case}. 
As a result, $V_1$ now passes all public tests and improves hidden-test performance from 10/15 to 13/15. 
However, $V_1$ is still not fully correct. 
It keeps the condition \texttt{c > 0}, which silently excludes 0 even though the problem statement explicitly allows non-positive answers. 
This remaining error is not exposed by the public tests, which motivates the need for Stage~C.

Stage~C is the key step. 
The Auditor reads the current code, notices the guard \texttt{c > 0}, and compares it against the specification that the answer is ``not necessarily positive.'' 
Based on this mismatch, it constructs a targeted corrective test with input \texttt{1 1} and the excluded set \texttt{\{1\}}. 
For this input, the nearest valid integers are 0 and 2, and the correct answer is 0 because ties should be broken by choosing the smaller one. 
The Auditor-generated test, therefore, reveals a hidden failure: $V_1$ outputs 2 instead of 0. 
This example highlights an important advantage of \textsc{FixAudit}: the Auditor does not rely on blind random testing, but instead uses the current code and the problem specification together to expose a failure mode that public tests miss.

In Stage~D, the Fixer uses this targeted failure signal to produce the final solution $V_2$. 
The change is minimal: \texttt{c > 0} becomes \texttt{c >= 0}. 
This single-character refinement is enough to fix the remaining bug and raise hidden-test performance from 13/15 to 15/15. 
The case study, therefore, illustrates the full value of our framework. 
Stage~B produces a plausible but incomplete repair based on the public test failure. 
Stage~C uncovers the hidden bug with a targeted test based on the current code and the problem specification. 
Stage~D then refines the code using this Auditor-generated test, instead of rewriting the whole program from scratch.

\section{Related Work}
\subsection{Prompt-Based and Agent-Based Code Generation}

Prompt-based methods improve code generation by revising the prompt before or after coding.
Post-generation techniques~\cite{selfdebugging,self_edit,selfrepair,reflexion,ldb_debug} use execution feedback to revise generated code after an initial attempt.
For example, Self-Repair~\cite{selfrepair} feeds failed-test feedback back into the prompt and asks the model to fix the current code.
Requirement-oriented methods~\cite{clarifygpt,mufix,requirements2code,intention2code} improve generation by clarifying or realigning the specification before (re)generating code.
For instance, ClarifyGPT~\cite{clarifygpt} asks clarifying questions to resolve ambiguities in the problem description before generating code.
Planning-based approaches~\cite{selfplanning,planningcode} add an explicit planning step before coding to decompose the problem into smaller steps.

Agent-based methods decompose the coding process into specialized roles such as planning, implementation, testing, and refinement~\cite{selfcollaboration,agentcoder,paircoder,mapcoder,soen101,alphacodium}.
AgentCoder~\cite{agentcoder} separates programming, test design, and test execution into distinct agents.
AlphaCodium~\cite{alphacodium} proposes a workflow-style pipeline that uses structured generation steps rather than one-shot prompting.
Among them, Specine~\cite{specine} is the most recent and strongest-performing framework in this category.
It extracts the model-perceived specification from a buggy candidate, rewrites the prompt to reduce ambiguity, and regenerates a new program from scratch.

\subsection{Reinforcement Learning for Code Generation and Unit-Test Generation}

Recent post-training methods include preference-optimization approaches such as CodeDPO~\cite{codedpo} and Focused-DPO~\cite{focuseddpo}, which align code models using self-generated preference data without execution feedback.
Another line uses RL with execution feedback as reward signals~\cite{le2022coderl,shojaee2023execution,liu2023rltf,dou2024stepcoder,gehring2024rlef}.
For example, CodeRL~\cite{le2022coderl} formulates code generation as policy optimization, and StepCoder~\cite{dou2024stepcoder} uses a curriculum over shorter subtasks to ease sparse-reward learning.
RLEF~\cite{gehring2024rlef} injects execution feedback into a multi-turn interaction before optimizing final correctness.
However, these methods optimize only the code generator and do not learn to produce new tests at inference time.

A separate line of work studies unit-test generation, including classical search-based methods~\cite{evosuite,randoop} and more recent neural or LLM-based approaches~\cite{focalcontext,chatunitest,testart,dynamic_unit_tests,ticoder}.
CodeT~\cite{codet} and LEVER~\cite{lever} show that generated tests or execution-based verifiers can help select better code candidates at inference time.
Steenhoek et al.~\cite{rl_unit_test} train a model to generate tests with automatic feedback for debugging purposes.
This line mainly focuses on test generation itself, rather than coupling a code-aware tester with a repair model in a closed debugging loop.

CURE~\cite{cure} unifies code generation and test generation through joint RL training, enabling test-time scaling~\cite{sstar}.
\textsc{FixAudit} extends this direction by having the Auditor read the candidate code for targeted test generation and coupling it with a Fixer for closed-loop repair.

\section{Threats to Validity}
\textbf{Internal validity} mainly lies in the fairness of comparisons.
To ensure that performance gains do not come from a larger inference budget, we align all frameworks to the same budget of 20 LLM invocations per problem (Section~4.4).
The framework baselines Specine and CURE use the same backbone model (Qwen2.5-Coder-7B-Instruct) and the same RL algorithm (DAPO) as \textsc{FixAudit} (Section~4.2).
To prevent data leakage, we apply strict $N$-gram filtering~\cite{brown2020language, codex_eval} to remove any overlap between training problems and downstream benchmarks (Section~4.3).

\textbf{External validity} mainly lies in the generalizability of our findings.
We evaluate \textsc{FixAudit} on two similarly sized backbones from the Qwen and Llama model families and focus on Python-based competitive programming problems.
Although both backbones show consistent gains, generalization to larger model scales and other programming languages remains to be verified.
To reduce benchmark-specific bias, we evaluate on four complementary benchmarks (APPS~\cite{apps}, CodeContests~\cite{alphacode}, xCodeEval~\cite{xcodeeval}, and LiveCodeBench v6~\cite{livecodebench}) and use two metrics that capture both strict correctness (Pass@1) and partial correctness (AvgPassRatio).

\section{Conclusion and Future Work}
In this paper, we proposed \textsc{FixAudit}, a framework that treats competitive code generation as a continuous, targeted test-and-repair cycle centered on the current candidate solution.
The framework trains two roles within a single shared model: a Fixer that performs localized repair while preserving correct logic, and an Auditor that reads the candidate code to generate tests targeting its specific bugs.
Through a four-stage training pipeline, including execution-aligned supervised fine-tuning followed by three stages of reinforcement learning, \textsc{FixAudit} first equips the model with execution reasoning and then progressively trains it for initial repair, test generation, and closed-loop refinement.
Across APPS, CodeContests, xCodeEval, and LiveCodeBench, \textsc{FixAudit} built on a 7B model achieves the strongest overall performance, surpassing the larger 32B model in the zero-shot setting and outperforming strong framework baselines on the same 7B backbone.

For future work, we plan to explore several directions.
First, we aim to extend \textsc{FixAudit} to larger backbone models and other programming languages beyond Python to test its generalizability.
Second, the current framework runs a fixed number of stages. An adaptive mechanism that decides when to stop repairing or when to generate additional tests could further improve efficiency.
Third, the current Auditor and Fixer are trained sequentially in separate RL stages.
We plan to explore joint or alternating training strategies that update both roles within the same RL loop, which may lead to stronger co-adaptation between the two agents.

\section{Data Availability}
We have released our code, data, and model at
\url{https://zenodo.org/records/21002377}.

\bibliographystyle{IEEEtran}
\bibliography{references}

@inproceedings{apps,
  author = {Dan Hendrycks and Steven Basart and Saurav Kadavath and Mantas Mazeika and Akul Arora and Ethan Guo and Collin Burns and Samir Puranik and Horace He and Dawn Song and Jacob Steinhardt},
  title = {Measuring Coding Challenge Competence With {APPS}},
  booktitle = {Thirty-fifth Conference on Neural Information Processing Systems Datasets and Benchmarks Track (Round 2)},
  year = {2021},
  url = {https://openreview.net/forum?id=sD93GOzH3i5}
}

@article{alphacode,
  author = {Yujia Li and David Choi and Junyoung Chung and Nate Kushman and Julian Schrittwieser and Remi Leblond and Tom Eccles and James Keeling and Felix Gimeno and Agustin Dal Lago and et al.},
  title = {Competition-Level Code Generation with {AlphaCode}},
  journal = {Science},
  volume = {378},
  number = {6624},
  pages = {1092--1097},
  year = {2022},
  doi = {10.1126/science.abq1158}
}

@inproceedings{xcodeeval,
  author = {Mohammad Abdullah Matin Khan and M. Saiful Bari and Xuan Long Do and Weishi Wang and Md Rizwan Parvez and Shafiq Joty},
  title = {{XCodeEval}: An Execution-Based Large Scale Multilingual Multitask Benchmark for Code Understanding, Generation, Translation and Retrieval},
  booktitle = {Proceedings of the 62nd Annual Meeting of the Association for Computational Linguistics (Volume 1: Long Papers)},
  pages = {6766--6805},
  year = {2024}
}

@inproceedings{selfrepair,
  author = {Theo X. Olausson and Jeevana Priya Inala and Chenglong Wang and Jianfeng Gao and Armando Solar-Lezama},
  title = {Is Self-Repair a Silver Bullet for Code Generation?},
  booktitle = {The Twelfth International Conference on Learning Representations},
  year = {2024},
  url = {https://openreview.net/forum?id=y0GJXRungR}
}

@inproceedings{specine,
  author = {Zhao Tian and Junjie Chen},
  title = {Aligning Requirement for Large Language Model's Code Generation},
  booktitle = {2026 IEEE/ACM 48th International Conference on Software Engineering (ICSE '26)},
  year = {2026},
  url = {https://arxiv.org/abs/2509.01313}
}

@article{cure,
  author = {Yinjie Wang and Ling Yang and Ye Tian and Ke Shen and Mengdi Wang},
  title = {Co-Evolving {LLM} Coder and Unit Tester via Reinforcement Learning},
  journal = {arXiv preprint arXiv:2506.03136},
  year = {2025},
  url = {https://arxiv.org/abs/2506.03136}
}

@article{grpo,
  author = {Zhihong Shao and Peiyi Wang and Qihao Zhu and Runxin Xu and Junxiao Song and Xiao Bi and Haowei Zhang and Mingchuan Zhang and Y. K. Li and Y. Wu and Daya Guo},
  title = {{DeepSeekMath}: Pushing the Limits of Mathematical Reasoning in Open Language Models},
  journal = {arXiv preprint arXiv:2402.03300},
  year = {2024},
  doi = {10.48550/arXiv.2402.03300}
}

@article{dapo,
  author = {Qiying Yu and Zheng Zhang and Ruofei Zhu and Yufeng Yuan and Xiaochen Zuo and Yu Yue and Weinan Dai and Tiantian Fan and Gaohong Liu and Lingjun Liu and Xin Liu and Haibin Lin and Zhiqi Lin and Bole Ma and Guangming Sheng and Yuxuan Tong and Chi Zhang and Mofan Zhang and Wang Zhang and Hang Zhu and Jinhua Zhu and Jiaze Chen and Jiangjie Chen and Chengyi Wang and Hongli Yu and Yuxuan Song and Xiangpeng Wei and Hao Zhou and Jingjing Liu and Wei-Ying Ma and Ya-Qin Zhang and Lin Yan and Mu Qiao and Yonghui Wu and Mingxuan Wang},
  title = {{DAPO}: An Open-Source {LLM} Reinforcement Learning System at Scale},
  journal = {arXiv preprint arXiv:2503.14476},
  year = {2025},
  doi = {10.48550/arXiv.2503.14476}
}

@article{qwen25coder,
  author = {Binyuan Hui and Jian Yang and Zeyu Cui and Jiaxi Yang and Dayiheng Liu and Lei Zhang and Tianyu Liu and Jiajun Zhang and Bowen Yu and Keming Lu and et al.},
  title = {Qwen2.5-Coder Technical Report},
  journal = {arXiv preprint arXiv:2409.12186},
  year = {2024},
  url = {https://arxiv.org/abs/2409.12186}
}

@article{requirements2code,
  author = {Bingyang Wei},
  title = {Requirements are All You Need: From Requirements to Code with {LLM}s},
  journal = {arXiv preprint arXiv:2406.10101},
  year = {2024},
  url = {https://arxiv.org/abs/2406.10101}
}

@inproceedings{intention2code,
  author = {Qi Guo and Xiaofei Xie and Shangqing Liu and Ming Hu and Xiaohong Li and Lei Bu},
  title = {Intention is All You Need: Refining Your Code from Your Intention},
  booktitle = {2025 IEEE/ACM 47th International Conference on Software Engineering (ICSE)},
  publisher = {IEEE Computer Society},
  year = {2025},
  doi = {10.1109/ICSE55347.2025.00191}
}

@article{selfplanning,
  author = {Xue Jiang and Yihong Dong and Lecheng Wang and Zheng Fang and Qiwei Shang and Ge Li and Zhi Jin and Wenpin Jiao},
  title = {Self-Planning Code Generation with Large Language Models},
  journal = {ACM Transactions on Software Engineering and Methodology},
  volume = {33},
  number = {7},
  pages = {1--30},
  year = {2024},
  doi = {10.1145/3672456}
}

@inproceedings{planningcode,
  author = {Shun Zhang and Zhenfang Chen and Yikang Shen and Mingyu Ding and Joshua B. Tenenbaum and Chuang Gan},
  title = {Planning with Large Language Models for Code Generation},
  booktitle = {The Eleventh International Conference on Learning Representations},
  year = {2023},
  url = {https://openreview.net/forum?id=Lr8cOOtYbfL}
}

@inproceedings{mapcoder,
  author = {Md. Ashraful Islam and Mohammed Eunus Ali and Md Rizwan Parvez},
  title = {MapCoder: Multi-Agent Code Generation for Competitive Problem Solving},
  booktitle = {Proceedings of the 62nd Annual Meeting of the Association for Computational Linguistics (Volume 1: Long Papers)},
  pages = {4912--4944},
  year = {2024},
  doi = {10.18653/v1/2024.acl-long.269}
}

@inproceedings{soen101,
  author = {Feng Lin and Dong Jae Kim and Tse-Hsun (Peter) Chen},
  title = {{SOEN-101}: Code Generation by Emulating Software Process Models Using Large Language Model Agents},
  booktitle = {2025 IEEE/ACM 47th International Conference on Software Engineering (ICSE)},
  year = {2025},
  doi = {10.1109/ICSE55347.2025.00140}
}

@article{codedpo,
  author = {Kechi Zhang and Ge Li and Yihong Dong and Jingjing Xu and Jun Zhang and Jing Su and Yongfei Liu and Zhi Jin},
  title = {{CodeDPO}: Aligning Code Models with Self Generated and Verified Source Code},
  journal = {arXiv preprint arXiv:2410.05605},
  year = {2024},
  url = {https://arxiv.org/abs/2410.05605}
}

@article{focuseddpo,
  author = {Kechi Zhang and Ge Li and Jia Li and Yihong Dong and Zhi Jin},
  title = {Focused-DPO: Enhancing Code Generation Through Focused Preference Optimization on Error-Prone Points},
  journal = {arXiv preprint arXiv:2502.11475},
  year = {2025},
  url = {https://arxiv.org/abs/2502.11475}
}

@article{dynamic_unit_tests,
  author = {Zeyao Ma and Xiaokang Zhang and Jing Zhang and Jifan Yu and Sijia Luo and Jie Tang},
  title = {Dynamic Scaling of Unit Tests for Code Reward Modeling},
  journal = {arXiv preprint arXiv:2501.01054},
  year = {2025},
  url = {https://arxiv.org/abs/2501.01054}
}

@inproceedings{evosuite,
  author = {Gordon Fraser and Andrea Arcuri},
  title = {EvoSuite: Automatic Test Suite Generation for Object-Oriented Software},
  booktitle = {Proceedings of the 19th ACM SIGSOFT Symposium and the 13th European Conference on Foundations of Software Engineering},
  pages = {416--419},
  year = {2011},
  doi = {10.1145/2025113.2025179}
}

@inproceedings{randoop,
  author = {Carlos Pacheco and Michael D. Ernst},
  title = {Randoop: Feedback-Directed Random Testing for Java},
  booktitle = {Companion to the 22nd ACM SIGPLAN Conference on Object-Oriented Programming Systems and Applications Companion},
  pages = {815--816},
  year = {2007},
  doi = {10.1145/1297846.1297902}
}

@article{focalcontext,
  author = {Michele Tufano and Dawn Drain and Alexey Svyatkovskiy and Shao Kun Deng and Neel Sundaresan},
  title = {Unit Test Case Generation with Transformers and Focal Context},
  journal = {arXiv preprint arXiv:2009.05617},
  year = {2020},
  url = {https://arxiv.org/abs/2009.05617}
}

@inproceedings{chatunitest,
  author = {Yinghao Chen and Zehao Hu and Chen Zhi and Junxiao Han and Shuiguang Deng and Jianwei Yin},
  title = {ChatUniTest: A Framework for {LLM}-Based Test Generation},
  booktitle = {Companion Proceedings of the 32nd ACM International Conference on the Foundations of Software Engineering},
  pages = {572--576},
  year = {2024},
  doi = {10.1145/3663529.3663801}
}

@article{testart,
  author = {Siqi Gu and Quanjun Zhang and Kecheng Li and Chunrong Fang and Fangyuan Tian and Liuchuan Zhu and Jianyi Zhou and Zhenyu Chen},
  title = {{TestART}: Improving {LLM}-Based Unit Testing via Co-Evolution of Automated Generation and Repair Iteration},
  journal = {arXiv preprint arXiv:2408.03095},
  year = {2024},
  url = {https://arxiv.org/abs/2408.03095}
}

@article{program_synthesis_llm,
  author = {Jacob Austin and Augustus Odena and Maxwell Nye and Maarten Bosma and Henryk Michalewski and David Dohan and Ellen Jiang and Carrie Cai and Michael Terry and Quoc Le and Charles Sutton},
  title = {Program Synthesis with Large Language Models},
  journal = {arXiv preprint arXiv:2108.07732},
  year = {2021},
  url = {https://arxiv.org/abs/2108.07732}
}

@article{codex_eval,
  author = {Mark Chen and Jerry Tworek and Heewoo Jun and Qiming Yuan and Henrique Ponde de Oliveira Pinto and Jared Kaplan and Harri Edwards and Yuri Burda and Nicholas Joseph and Greg Brockman and Alex Ray and Raul Puri and Gretchen Krueger and Girish Sastry and Amanda Askell and Pamela Mishkin and Jack Clark and Carroll Wainwright and Mira Murati and Dario Amodei and Sam McCandlish and Ilya Sutskever and Wojciech Zaremba},
  title = {Evaluating Large Language Models Trained on Code},
  journal = {arXiv preprint arXiv:2107.03374},
  year = {2021},
  url = {https://arxiv.org/abs/2107.03374}
}

@article{deepseekcoder,
  author = {Daya Guo and Qihao Zhu and Dejian Yang and Zhenda Xie and Kai Dong and Wentao Zhang and Guanting Chen and Xiao Bi and Yu Wu and Y. K. Li and et al.},
  title = {{DeepSeek-Coder}: When the Large Language Model Meets Programming---The Rise of Code Intelligence},
  journal = {arXiv preprint arXiv:2401.14196},
  year = {2024},
  url = {https://arxiv.org/abs/2401.14196}
}

@article{rigorous_eval,
  author = {Jiawei Liu and Chunqiu Steven Xia and Yuyao Wang and Lingming Zhang},
  title = {Is Your Code Generated by {ChatGPT} Really Correct? Rigorous Evaluation of Large Language Models for Code Generation},
  journal = {Advances in Neural Information Processing Systems},
  volume = {36},
  pages = {21558--21572},
  year = {2023},
  doi = {10.52202/075280-0943}
}

@article{livecodebench,
  author = {Naman Jain and King Han and Alex Gu and Wen-Ding Li and Fanjia Yan and Tianjun Zhang and Sida I. Wang and Armando Solar-Lezama and Koushik Sen and Ion Stoica},
  title = {{LiveCodeBench}: Holistic and Contamination-Free Evaluation of Large Language Models for Code},
  journal = {arXiv preprint arXiv:2403.07974},
  year = {2024},
  url = {https://arxiv.org/abs/2403.07974}
}

@article{selfdebugging,
  author = {Xinyun Chen and Maxwell Lin and Nathanael Sch{\"a}rli and Denny Zhou},
  title = {Teaching Large Language Models to Self-Debug},
  journal = {The Twelfth International Conference on Learning Representations},
  year = {2024},
  url = {https://openreview.net/forum?id=KuPixIqPiq}
}

@article{selfcollaboration,
  author = {Yihong Dong and Xue Jiang and Zhi Jin and Ge Li},
  title = {Self-Collaboration Code Generation via {ChatGPT}},
  journal = {ACM Transactions on Software Engineering and Methodology},
  volume = {33},
  number = {7},
  pages = {1--38},
  year = {2024},
  doi = {10.1145/3672459}
}

@inproceedings{paircoder,
  author = {Huan Zhang and Wei Cheng and Yuhan Wu and Wei Hu},
  title = {A Pair Programming Framework for Code Generation via Multi-Plan Exploration and Feedback-Driven Refinement},
  booktitle = {Proceedings of the 39th IEEE/ACM International Conference on Automated Software Engineering},
  pages = {1319--1331},
  year = {2024},
  doi = {10.1145/3691620.3695506}
}

@article{clarifygpt,
  author = {Fangwen Mu and Lin Shi and Song Wang and Zhuohao Yu and Binquan Zhang and ChenXue Wang and Shichao Liu and Qing Wang},
  title = {{ClarifyGPT}: A Framework for Enhancing {LLM}-Based Code Generation via Requirements Clarification},
  journal = {Proceedings of the ACM on Software Engineering},
  volume = {1},
  number = {FSE},
  pages = {2332--2354},
  year = {2024},
  doi = {10.1145/3660810}
}

@inproceedings{mufix,
  author = {Zhao Tian and Junjie Chen and Xiangyu Zhang},
  title = {Fixing Large Language Models' Specification Misunderstanding for Better Code Generation},
  booktitle = {2025 IEEE/ACM 47th International Conference on Software Engineering (ICSE)},
  year = {2025},
  doi = {10.1109/ICSE55347.2025.00108}
}

@article{codet,
  author = {Bei Chen and Fengji Zhang and Anh Nguyen and Daoguang Zan and Zeqi Lin and Jian-Guang Lou and Weizhu Chen},
  title = {{CodeT}: Code Generation with Generated Tests},
  journal = {The Eleventh International Conference on Learning Representations},
  year = {2023},
  url = {https://openreview.net/forum?id=ktrw68Cmu9c}
}

@article{sstar,
  author = {Dacheng Li and Shiyi Cao and Chengkun Cao and Xiuyu Li and Shangyin Tan and Kurt Keutzer and Jiarong Xing and Joseph E. Gonzalez and Ion Stoica},
  title = {{S*}: Test Time Scaling for Code Generation},
  journal = {arXiv preprint arXiv:2502.14382},
  year = {2025},
  url = {https://arxiv.org/abs/2502.14382}
}

@article{semcoder,
  author = {Yangruibo Ding and Jinjun Peng and Marcus Min and Gail Kaiser and Junfeng Yang and Baishakhi Ray},
  title = {{SemCoder}: Training Code Language Models with Comprehensive Semantics Reasoning},
  journal = {Advances in Neural Information Processing Systems},
  volume = {37},
  pages = {60275--60308},
  year = {2024},
  doi = {10.52202/079017-1927}
}

@inproceedings{reflexion,
  title={Reflexion: Language agents with verbal reinforcement learning},
  author={Shinn, Noah and Cassano, Federico and Gopinath, Ashwin and Narasimhan, Karthik and Yao, Shunyu},
  booktitle={Advances in Neural Information Processing Systems},
  volume={36},
  year={2023},
  doi={10.52202/075280-0377}
}

@inproceedings{ldb_debug,
  title={Debug like a Human: A Large Language Model Debugger via Verifying Runtime Execution Step by Step},
  author={Zhong, Li and Wang, Zilong and Shang, Jingbo},
  booktitle={Findings of the Association for Computational Linguistics: ACL 2024},
  year={2024},
  doi={10.18653/v1/2024.findings-acl.49}
}

@inproceedings{rl_unit_test,
  title={Reinforcement Learning from Automatic Feedback for High-Quality Unit Test Generation},
  author={Steenhoek, Benjamin and Tufano, Michele and Sundaresan, Neel and Svyatkovskiy, Alexey},
  booktitle={2025 IEEE/ACM International Workshop on Deep Learning for Testing and Testing for Deep Learning (DeepTest)},
  year={2025},
  doi={10.1109/DeepTest66595.2025.00011}
}

@article{ticoder,
  title={{LLM}-Based Test-Driven Interactive Code Generation: User Study and Empirical Evaluation},
  author={Fakhoury, Sarah and Naik, Aaditya and Sakkas, Georgios and Chakraborty, Saikat and Lahiri, Shuvendu K.},
  journal={IEEE Transactions on Software Engineering},
  volume={50},
  number={9},
  pages={2254--2268},
  year={2024},
  doi={10.1109/TSE.2024.3428972}
}

@inproceedings{lever,
  title={{LEVER}: Learning to Verify Language-to-Code Generation with Execution},
  author={Ni, Ansong and Iyer, Srini and Radev, Dragomir and Stoyanov, Veselin and Yih, Wen-Tau and Wang, Sida and Lin, Xi Victoria},
  booktitle={Proceedings of the 40th International Conference on Machine Learning},
  pages={26106--26128},
  year={2023},
  url={https://proceedings.mlr.press/v202/ni23b.html}
}

@inproceedings{self_edit,
  title={{Self-Edit}: Fault-Aware Code Editor for Code Generation},
  author={Zhang, Kechi and Li, Zhuo and Li, Jia and Li, Ge and Jin, Zhi},
  booktitle={Proceedings of the 61st Annual Meeting of the Association for Computational Linguistics},
  year={2023},
  doi={10.18653/v1/2023.acl-long.45}
}

@article{ppo,
  title={Proximal policy optimization algorithms},
  author={Schulman, John and Wolski, Filip and Dhariwal, Prafulla and Radford, Alec and Klimov, Oleg},
  journal={arXiv preprint arXiv:1707.06347},
  year={2017},
  url={https://arxiv.org/abs/1707.06347}
}

@article{le2022coderl,
  title={Coderl: Mastering code generation through pretrained models and deep reinforcement learning},
  author={Le, Hung and Wang, Yue and Gotmare, Akhilesh Deepak and Savarese, Silvio and Hoi, Steven Chu Hong},
  journal={Advances in Neural Information Processing Systems},
  volume={35},
  pages={21314--21328},
  year={2022},
  doi={10.52202/068431-1549}
}

@article{shojaee2023execution,
  title={Execution-based code generation using deep reinforcement learning},
  author={Shojaee, Parshin and Jain, Aneesh and Tipirneni, Sindhu and Reddy, Chandan K},
  journal={arXiv preprint arXiv:2301.13816},
  year={2023},
  url={https://arxiv.org/abs/2301.13816}
}

@article{liu2023rltf,
  title={Rltf: Reinforcement learning from unit test feedback},
  author={Liu, Jiate and Zhu, Yiqin and Xiao, Kaiwen and Fu, Qiang and Han, Xiao and Yang, Wei and Ye, Deheng},
  journal={arXiv preprint arXiv:2307.04349},
  year={2023},
  url={https://arxiv.org/abs/2307.04349}
}

@inproceedings{dou2024stepcoder,
  title={Stepcoder: improving code generation with reinforcement learning from compiler feedback},
  author={Dou, Shihan and Liu, Yan and Jia, Haoxiang and Zhou, Enyu and Xiong, Limao and Shan, Junjie and Huang, Caishuang and Wang, Xiao and Fan, Xiaoran and Xi, Zhiheng and others},
  booktitle={Proceedings of the 62nd Annual Meeting of the Association for Computational Linguistics (Volume 1: Long Papers)},
  pages={4571--4585},
  year={2024},
  doi={10.18653/v1/2024.acl-long.251}
}

@article{gehring2024rlef,
  title={Rlef: Grounding code llms in execution feedback with reinforcement learning},
  author={Gehring, Jonas and Zheng, Kunhao and Copet, Jade and Mella, Vegard and Carbonneaux, Quentin and Cohen, Taco and Synnaeve, Gabriel},
  journal={arXiv preprint arXiv:2410.02089},
  year={2024},
  url={https://arxiv.org/abs/2410.02089}
}

@article{agentcoder,
  author = {Dong Huang and Jie M. Zhang and Michael Luck and Qingwen Bu and Yuhao Qing and Heming Cui},
  title = {{AgentCoder}: Multi-Agent-Based Code Generation with Iterative Testing and Optimisation},
  journal = {arXiv preprint arXiv:2312.13010},
  year = {2023},
  url = {https://arxiv.org/abs/2312.13010}
}

@article{alphacodium,
  author = {Tal Ridnik and Dedy Kredo and Itamar Friedman},
  title = {Code Generation with {AlphaCodium}: From Prompt Engineering to Flow Engineering},
  journal = {arXiv preprint arXiv:2401.08500},
  year = {2024},
  url = {https://arxiv.org/abs/2401.08500}
}

@misc{qwq32b,
    title = {QwQ-32B: Embracing the Power of Reinforcement Learning},
    url = {https://qwenlm.github.io/blog/qwq-32b/},
    author = {Qwen Team},
    month = {March},
    year = {2025}
}

@article{qwen2.5,
      title={Qwen2.5 Technical Report}, 
      author={An Yang and Baosong Yang and Beichen Zhang and Binyuan Hui and Bo Zheng and Bowen Yu and Chengyuan Li and Dayiheng Liu and Fei Huang and Haoran Wei and Huan Lin and Jian Yang and Jianhong Tu and Jianwei Zhang and Jianxin Yang and Jiaxi Yang and Jingren Zhou and Junyang Lin and Kai Dang and Keming Lu and Keqin Bao and Kexin Yang and Le Yu and Mei Li and Mingfeng Xue and Pei Zhang and Qin Zhu and Rui Men and Runji Lin and Tianhao Li and Tianyi Tang and Tingyu Xia and Xingzhang Ren and Xuancheng Ren and Yang Fan and Yang Su and Yichang Zhang and Yu Wan and Yuqiong Liu and Zeyu Cui and Zhenru Zhang and Zihan Qiu},
      journal={arXiv preprint arXiv:2412.15115},
      year={2024},
      url={https://arxiv.org/abs/2412.15115}
}

@article{taco,
  title={{TACO}: Topics in Algorithmic Code Generation Dataset},
  author={Rongao Li and Jie Fu and Bo-Wen Zhang and Tao Huang and Zhihong Sun and Chen Lyu and Guang Liu and Zhi Jin and Ge Li},
  journal={arXiv preprint arXiv:2312.14852},
  year={2023},
  url={https://arxiv.org/abs/2312.14852}
}

@article{brown2020language,
  title={Language models are few-shot learners},
  author={Brown, Tom and Mann, Benjamin and Ryder, Nick and Subbiah, Melanie and Kaplan, Jared D and Dhariwal, Prafulla and Neelakantan, Arvind and Shyam, Pranav and Sastry, Girish and Askell, Amanda and others},
  journal={Advances in neural information processing systems},
  volume={33},
  pages={1877--1901},
  year={2020},
  url={https://proceedings.neurips.cc/paper/2020/hash/1457c0d6bfcb4967418bfb8ac142f64a-Abstract.html}
}

@misc{verl,
  title={veRL: Volcano Engine Reinforcement Learning for LLM},
  author={Sheng, Jianjian and others},
  year={2024},
  url={https://github.com/volcengine/verl}
}

@misc{gpt4o,
  title={Hello GPT-4o},
  author={{OpenAI}},
  year={2024},
  url={https://openai.com/index/hello-gpt-4o/}
}

@article{gu2024cruxeval,
  title={Cruxeval: A benchmark for code reasoning, understanding and execution},
  author={Gu, Alex and Rozi{\`e}re, Baptiste and Leather, Hugh and Solar-Lezama, Armando and Synnaeve, Gabriel and Wang, Sida I},
  journal={arXiv preprint arXiv:2401.03065},
  year={2024},
  url={https://arxiv.org/abs/2401.03065}
}

@inproceedings{chen2025reasoning,
  title={Reasoning runtime behavior of a program with llm: How far are we?},
  author={Chen, Junkai and Pan, Zhiyuan and Hu, Xing and Li, Zhenhao and Li, Ge and Xia, Xin},
  booktitle={2025 IEEE/ACM 47th International Conference on Software Engineering (ICSE)},
  pages={1869--1881},
  year={2025},
  organization={IEEE},
  doi={10.1109/ICSE55347.2025.00012}
}

@inproceedings{muennighoff2025s1,
  title={s1: Simple test-time scaling},
  author={Muennighoff, Niklas and Yang, Zitong and Shi, Weijia and Li, Xiang Lisa and Fei-Fei, Li and Hajishirzi, Hannaneh and Zettlemoyer, Luke and Liang, Percy and Cand{\`e}s, Emmanuel and Hashimoto, Tatsunori B},
  booktitle={Proceedings of the 2025 Conference on Empirical Methods in Natural Language Processing},
  pages={20286--20332},
  year={2025},
  doi={10.18653/v1/2025.emnlp-main.1025}
}

\end{document}